\documentclass{aa}
\usepackage{graphicx}

\newcommand{\beq}{\begin{equation}}
\newcommand{\eeq}{\end{equation}}
\newcommand{\beqa}{\begin{eqnarray}}
\newcommand{\eeqa}{\end{eqnarray}}

\begin{document}

\title{Anomalous fluctuations in observations of Q0957+561~A,B:
smoking gun of a cosmic string?}
\titlerunning{Anomalous fluctuations in observations of Q0957+561~A,B.}

\author{R. Schild \inst{1}
\and I. S. Masnyak \inst{2}
\and B. I. Hnatyk \inst{2}
\and V. I. Zhdanov \inst{2}}
\authorrunning{R. Schild et al.}

\mail{rschild@cfa.harvard.edu}

\institute{ Harvard-Smithsonian Center for Astrophysics, 19, 60 Garden Street,
Cambridge, MA 02138, U.S.A.
\and Astronomical Observatory of Kyiv Taras Shevchenko National
University, 3 Observatorna str., 04053  Kyiv, Ukraine}

\abstract{ We report  the detection of anomalous brightness fluctuations in the multiple
image Q0957 + 561~A,B gravitational lens system, and consider whether such anomalies
have a plausible interpretation within the framework of cosmic string theory. We
study a simple model of gravitational lensing by an asymmetric rotating string. An explicit
form of the lens equation is obtained and approximate relations for magnification
are derived. We show that such a model with typical parameters of the GUT string
can quantitatively reproduce the observed pattern of brightness fluctuations.
On the other hand explanation involving a binary star system as an alternative
cause requires an unacceptably large massive object at a small distance. We also
discuss possible observational manifestations of cosmic strings within our lens model.
}


\maketitle

\section{Introduction}

Recent observations of the Q0957+561 A,B gravitational lens system show unexpected
synchronous (without the expected time delay) fluctuations of brightness of the two
quasar images. An ordinary binary star system, which might theoretically explain
such fluctuations, would be too massive and close to us, i.e., would be clearly
visible, which is not the case. Therefore, we attempt to explain these data by
lensing on a cosmic string, particularly on a loop of string. The existence of
cosmic strings is predicted by particle physics (Vilenkin \& Shellard \cite{VilShe94}) and
gravitational lensing effects are a promising signature of these astrophysical
objects. Sazhin et al. (\cite{Sazetal03}) claimed the detection of the first case of
cosmic string
lensing. We demonstrate here another possible signature of a string: microlensing by
oscillating loops of cosmic strings, which results in quasiperiodic fluctuations of
the observed brightness of the source. In section~\ref{Obs} we discuss the observational
data, and in section~\ref{Gra} a quantitative model of string lensing is elaborated.
In section~\ref{Exp} an explanation of the observational data is presented, with
discussion and conclusions in section~\ref{Dis}.

\section {Observations of an anomaly in the Q0957+561 A,B gravitational
lens system}
\label{Obs}

The Q0957 system was the first discovered multiple image gravitational lens system,
and already at the time of its discovery in 1979 (Walsh et al. \cite{Wal}) it was understood
that it was extremely important to astrophysics. Measurement of the time delay between
fluctuations in the two known images would allow determination of the Hubble constant from
simple theory, independent of uncertainty in local distance estimates for the supernovae
and Cepheid variable stars (Refsdal \cite{Ref64}). Thus from the time of discovery, monitoring
of the brightness of the two images, separated by 6 arcsec, was undertaken so that the
quasar's intrinsic brightness fluctuations could be recognized in the two images
separately, and a time delay measured.

With the Schild and Cholfin (\cite{SC86}) measurement of time delay (including numerous subsequent
refinements; see Colley et al (\cite{Col03}) for a summary) it was soon recognized that there were
differences between the time delay corrected brightness curves, although the basic pattern
could be easily recognized. The differences were attributed to microlensing by individual
massive objects, presumably stars, in the lens galaxy (Schild \& Smith \cite{SS91}).
The prospect
that such microlensing might allow detection of any baryonic missing mass objects justified
intensive monitoring campaigns, and in the 24 years since discovery the source has been
consistently observed on more than 1500 nights.

Such monitoring reveals two principal components in the quasar's brightness fluctuations:
a component due to intrinsic quasar brightness fluctuations, first seen in image A and then
seen 417.1 days later in image B, and a microlensing component arising in only one image
component due to individual stars along the A or B image line of sight.

\begin{figure}[!]
\resizebox{\hsize}{!}{\includegraphics{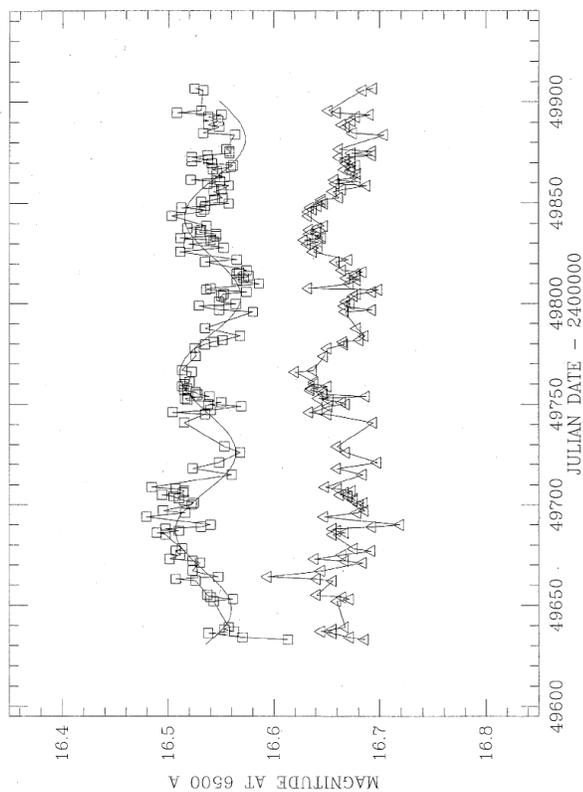}}
\caption {Brightness of the two quasar images displayed with no correction
for gravitational lens time delay. The brightness of quasar image A (upper
record with square data markers), has been fitted with a sine curve having
0.04 magnitude amplitude. The lower record, with triangular markers, appears
to have the same sinusoidal brightness curve with zero lag, even though at
most epochs data for the gravitationally lensed images show a lag of 417
days.}
\label{fig1}
\end{figure}

\begin{figure}[!]
\resizebox{\hsize}{!}{\includegraphics{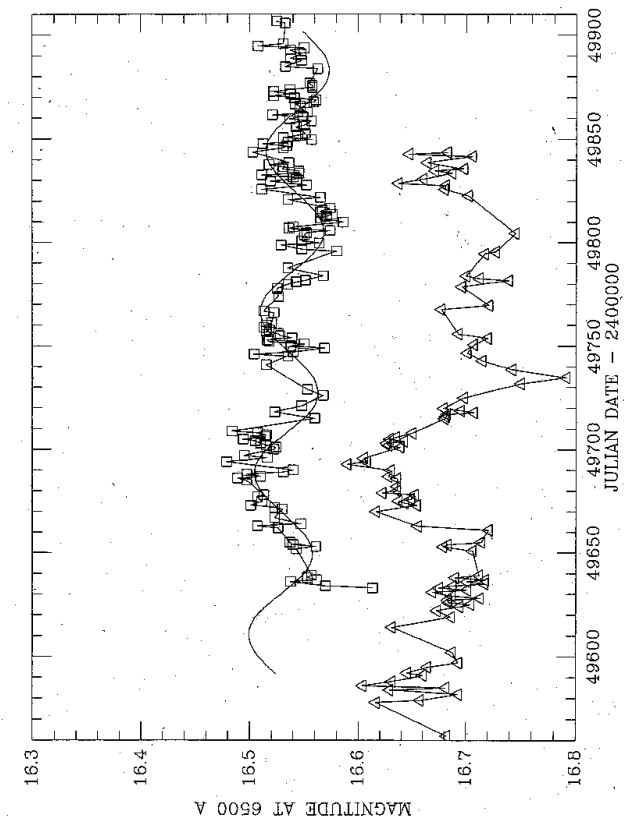}}
\caption {Quasar brightness displayed for measured time delay. The upper
record shows the same data and fit for image A as displayed in Fig.~\ref{fig1}.
The lower data markers (triangles) are the brightness measurements for
image B measured 417 days (the gravitational lens time delay) later, but
with 417 subtracted from the Julian dates for plotting. If the image A
brightness fluctuations are intrinsic to the quasar, they should be seen
also in image B 417 days later, but the two are seen not to match as well
as the 0 lag comparison in Fig.~\ref{fig1}.}
\label{fig2}
\end{figure}

\begin{figure}[!]
\resizebox{\hsize}{!}{\includegraphics{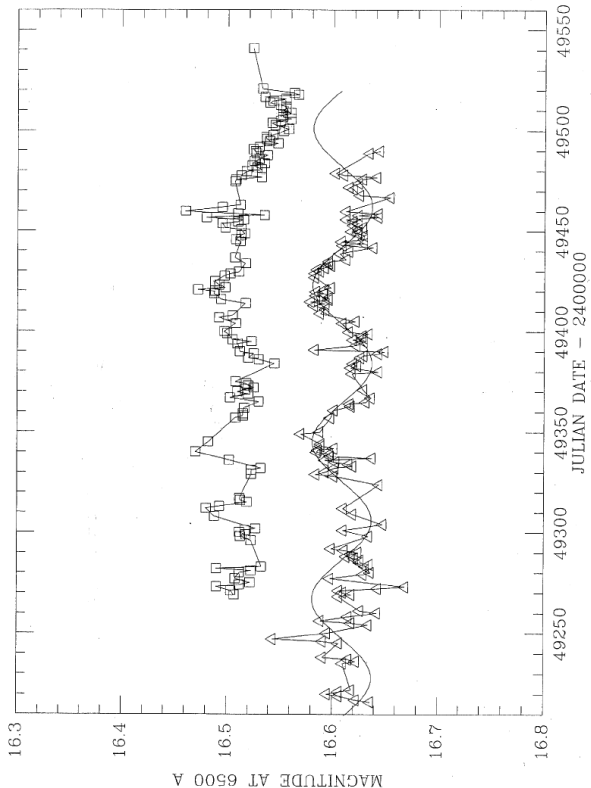}}
\caption {Quasar brightness with 417-day lag for the opposite image pair.
The lower record is the Fig.~\ref{fig1} data for image B, with the data for image A
measured 417 days previously. The agreement is seen to be poor, especially
near the end of the observational period, even though the theory of
gravitational lenses shows that the time delay must produce agreement for
417 days. If the measured sinusoidal oscillation seen in both images A and
B (Fig.~\ref{fig1}) is a chance coincidence of two quasar oscillations separated by
417 days, there must be agreement for both A data with corrected B data
(Fig.~\ref{fig2}) and B data with corrected A data (this figure). Comparing Fig.~\ref{fig1}
with Figs.~\ref{fig2} and \ref{fig3} shows best agreement for 0 lag, contrary to gravitational
lens time delay theory. }
\label{fig3}
\end{figure}

We now illustrate what appears to be a third component of quasar brightness fluctuations,
seen in the combination of Figs.~\ref{fig1}, \ref{fig2} and \ref{fig3}. Here we plot
the measured brightnesses of the two quasar images as measured during the 1994-1995
and 1995-1996 seasons. In Fig.~\ref{fig1} no
correction has been made for time delay; the plotted brightness measurements in magnitudes
are shown for the Julian dates of observation. The plotted symbols are the size of the
typical $1-\sigma$ error bars previously established for this data set (.006 and .007 mag
for images A and B). In Fig.~\ref{fig1} a sine curve has been fitted to the A image data
but not B image to allow the eye to judge whether there appears to be a repeating pattern of
fluctuations for 0 lag.

Fig.~\ref{fig1} shows the unexpected result that a short-duration oscillation of $4\%$ amplitude
and periodicity of approximately $T \approx 100$ days was seen for approximately 400 days. The
amplitude
of these fluctuations is well above the known errors of the measurements. The error estimates
originally attributed to these data by Schild (\cite{ST95}) have been confirmed from subsequent
analysis by Colley and Schild (\cite{Col99}), and the entire data set has subsequently been
re-reduced by Ovaldsen et al. (\cite{Ova03}), who also call attention to the observed
correlation for 0 lag.

The correlation for 0 lag is not perfect, as would be expected, since several processes are
simultaneously causing brightness fluctuations. Microlensing can impose a random pattern of
fluctuations with durations ranging from 1 day to decades. An example of an event with 0.01
mag amplitude and 12 hour duration has been given by Colley and Schild (\cite{CS03}).
A wavelet analysis of the long brightness record by Schild (\cite{RS99}) shows that
events on time scales of
1 and 60 days have typical amplitudes of 0.01 and 0.08 magnitudes, respectively. Yet the
fluctuation pattern is sometimes seen to effectively stop, as reported in Colley et al.
(\cite{Col03}).

If the observed fluctuations are caused by random microlensing events, it would be
unexpected for them to be apparently in phase. We have not yet devised a statistical test
defining the significance level or error limits on simultaneity because the basic statistical
process is non-Gaussian and has not yet been simulated. Moreover, any statistical analysis
cannot be perfect in the presence of the usual stochastic microlensing variation taking into
account the limited time interval where the anomalous effect has been observed. If the
fluctuations are intrinsic to the quasar, and seen simultaneously by some highly improbable
coincidence, they must be seen in the observations of the preceding and following years, as
illustrated in Figs.~\ref{fig2} and \ref{fig3}. Thus we show in Fig.~\ref{fig2} that if
the B data of Fig.~\ref{fig1} are compared
to A image data measured 417 days previously, the fluctuations are probably not seen.
Similarly we illustrate in Fig.~\ref{fig3} that if the A image pattern from Fig.~\ref{fig1}
is compared to B image 417 days later, the pattern is again not seen. If the brightness
fluctuations are
intrinsic to the quasar and seen simultaneously as in Fig.~\ref{fig1} by chance,
they must also be
seen in the lagged data for the opposite quasar image; thus they would be seen in both
Fig.~\ref{fig2} and Fig.~\ref{fig3}.

The importance of these observations relates to the fact that there should be no causal
connection between brightness fluctuations seen simultaneously. If the fluctuations were due
to the quasar's intrinsic brightness changes, then they should be seen at the measured time
delay, which is  $417.09 \pm 0.07$ days (Colley et al. \cite{Col03}). If they were produced in
proximity to the lens galaxy at redshift 0.37 (the quasar redshift is 1.41) they should
similarly be seen with a large time delay. Only fluctuations produced locally (i.e., between
the lens galaxy and the observer but close to the observer) can be observed to be simultaneous.

This problem is even more serious because of the relatively large separation of the two quasar
images, 6.2~arcsec on the sky. Supposition that the above oscillations are induced by orbiting
of binary stars leads to anomalously large masses of the components, as shown in
subsection~\ref{expl:osc}. Therefore we consider below the possibility that the oscillations
are due to time variations of the gravitational field of a cosmic string.

\section{Gravitational lensing by a cosmic string}
\label{Gra}

\subsection{Cosmic string characteristics}

Cosmic strings are linear defects that could be formed at a symmetry breaking phase
transition in the early Universe (Vilenkin \& Shellard \cite{VilShe94}).

A horizon-sized volume at any cosmological time $t$ should contain a few long strings
stretching across the volume as well as a large number of small closed loops. At the moment
of creation $t$, typical loop length is $l \sim \alpha c t$, i.e., about $\alpha $ of
horizon size $ct$. During the string evolution, loops constitute some fixed part of total
string network; this scaling  results in the following loop number density
\beq
n_l(t)\sim \alpha^{-1}(ct)^{-3}.
\label{n}
\eeq
Typically $\alpha$ is determined by the gravitational back-reaction, so that
$\alpha\sim k_\mathrm{g} G\mu/c^2=k_\mathrm{g}\epsilon $, where  $k_\mathrm{g}\sim 50$
is a numerical coefficient, $G$ is gravitational constant, $c$ is the speed of light, $\mu\sim \eta^2/\hbar c^3$
is the mass per unit length of string, $\hbar$ is the Planck constant, and $\eta$ is
the symmetry breaking scale of strings. For a Grand Unified Theory (GUT) string
$\eta_\mathrm{GUT}\sim 10^{25}$~eV and $\mu_\mathrm{GUT}\sim 10^{22}$~g/cm,
$\epsilon_\mathrm{GUT} \sim 10^{-6}$ and $\alpha_\mathrm{GUT} \sim 10^{-4}$.

The loops oscillate and lose their energy, mostly by gravitational radiation. For a loop
of length $l$, the oscillation period is $T_l=l/2c$ and the lifetime determined by
gravitational radiation losses is $\tau_l\sim l/c(k_\mathrm{g} G\mu/c^2)$.

Gravitational lensing by cosmic strings has been considered by many authors (see references
in Vilenkin \& Shellard (\cite{VilShe94}) and de Laix \& Vachaspati (\cite{Laix})).
Straight cosmic strings
have a distinctive feature: they  produce two identical images. However they cannot explain
the oscillatory character of our data. Therefore we consider gravitational action of cosmic
string loops, which cause effects similar to ordinary oscillating systems (binary stars and
others), but are more massive and move with relativistic speeds.

\subsection{Lensing by oscillating  loops}

De Laix \& Vachaspati (\cite{Laix}) considered in detail, lensing by
cosmic string loops. Here we use their approach for the
interpretation of the observed oscillations. In the simplest
idealized case of a circular loop, with oscillations reduced to
variations of loop radius, they find that the image brightness of
a point source will not oscillate if the loop does not overlap the
source. Consequently we should take an asymmetric loop to explain
the observed oscillations. We consider a
maximally asymmetrical string configuration in the form of a rotating
double line segment of length $2R$ lying transverse to the line of
sight and having coordinates:
\beqa
x_1^{str}&=&R\cos{(ct/R)}\sin{\sigma},\nonumber\\
x_2^{str}&=&R\sin{(ct/R)}\sin{\sigma},\label{coord}\\
x_3^{str}&=&0.\nonumber
\eeqa
This is a particular solution from the family of known solutions to string equations
(Turok \cite{Turok}; Vilenkin \& Shellard \cite{VilShe94}).
Here axis $x_3$ is directed to observer,
$x_1^{str}$, $x_2^{str}$ are coordinates of the loop in the lens
plane, $\sigma$ indicates position on the string and varies from
$0$ to $2\pi$ and $t$ is the time. This configuration is a limiting case of
asymmetric loop; in a more realistic case a loop will have an
ellipsoidal form with large eccentricity.

The lens equations can be obtained from general result of de Laix \& Vachaspati (\cite{Laix}).
After some calculations taking into account the particular solution (\ref{coord})
we have:
\beqa
&&\frac{D_l}{D_s}\tilde{y}_{1}=\tilde{x}_{1}-q_\mathrm{s}R^2\;\mathrm{sgn}\,(\tilde{x}_1)\times\nonumber\\
&&\times\sqrt{\frac{\sqrt{\big(R^2+\tilde{x}_{1}^{2}+\tilde{x}_{2}^{2}\big)^2
-4R^2\tilde{x}_{1}^{2}}
-R^2-\tilde{x}_{2}^{2}+\tilde{x}_{1}^{2}}
{2\big(\big(R^2+\tilde{x}_{1}^{2}+\tilde{x}_{2}^{2}\big)^2
-4R^2\tilde{x}_{1}^{2}\big)}},
\label{lenseq1}
\\
&&\frac{D_l}{D_s}\tilde{y}_{2}=\tilde{x}_{2}-q_\mathrm{s}R^2\;\mathrm{sgn}\,(\tilde{x}_2)\times\nonumber\\
&&\times\sqrt{\frac{\sqrt{\big(R^2+\tilde{x}_{1}^{2}+\tilde{x}_{2}^{2}\big)^2
-4R^2\tilde{x}_{1}^{2}} +R^2+\tilde{x}_{2}^{2}-\tilde{x}_{1}^{2}}
{2\big(\big(R^2+\tilde{x}_{1}^{2}+\tilde{x}_{2}^{2}\big)^2 -4R^2
\tilde{x}_{1}^{2}\big)}},
\label{lenseq2}
\eeqa
where
$q_\mathrm{s}=8\pi\frac{G\mu}{c^2}\frac{D_{ls}D_l}{D_s R}$
($D_s$, $D_l$ and $D_{ls}$ are distances from us to source
plane, to lens plane and from source to lens plane
respectively), $\tilde{x}_1={x}_1\cos{(ct/
R)}+{x}_2\sin{(ct/ R)}$, $\tilde{x}_2={x}_2\cos{(ct/
R)}-{x}_1\sin{(ct/ R)}$ (where $x_1$, $x_2$ are coordinates
in the lens plane) and $\tilde{y}_1={y}_1\cos{(ct/
R)}+{y}_2\sin{(ct/ R)}$, $\tilde{y}_2={y}_2\cos{(ct/
R)}-{y}_1\sin{(ct/ R)}$ (where $y_1$, $y_2$ are coordinates in the
source plane). It may be shown that all the relations under the root signs
are non-negative.

Magnification of a point-like source by such a string is
\beqa
&&m=\Bigg|1-q_\mathrm{s}^2R^4 ({x}_1^2+{x}_2^2)\Big/\Big((R^2+{x}_1^2+{x}_2^2)^2-\nonumber\\
&&-4R^2({x}_1\cos{\frac{c\,t}{ R}} +{x}_2\sin{\frac{c\,t}{
R}})^2\Big)^{3/2}\Bigg|^{-1}.
\label{mag1}
\eeqa
If $q_\mathrm{s}\sim1$ and $R_y=R D_s/D_l\ll\rho=(y_1^2+y_2^2)^{1/2}$,
we can solve approximately the lens equations (\ref{lenseq1},\ref{lenseq2})
and obtain the magnification
\beqa
&&m=1+\frac{q_\mathrm{s}^2 R_y^4}{\rho^4}
-\frac{4q_\mathrm{s}^3 R_y^6}{\rho^6}
\nonumber\\
&&+\frac{3q_\mathrm{s}^2 R_y^6}{\rho^8}
\Big(({y}_1^2-{y}_2^2)\cos{\frac{2ct}{ R}}
+2{y}_1{y}_2\sin{\frac{2ct}{ R}}\Big).
\label{expan}
\eeqa
This yields amplitude of source brightness fluctuations as follows
\beqa
\Delta m\approx\frac{6
q_\mathrm{s}^2R_y^6}{\rho^6} \approx\frac{384\pi^2
G^2\mu^{2}\theta_{R}^4}{c^4\theta_I^6},
\label{magnif1}
\eeqa
where $\theta_I=\rho/D_s$ is the angular impact distance of
the line-of-sight with respect to the center of the loop,
$\theta_R=R/D_l$ is half of the visible angular size of the loop.

\subsection{Lensing by a binary system}

Now we consider for comparison the gravitational lensing by a binary system
of two equal point masses $M$ orbiting their center of mass with
the period $T$. Further, $r$ is half of the distance between the masses
and $\omega=\pi/T$.
The magnification of a point-like source by such lens system is (Schneider, Ehlers \&
Falco \cite{Schneider}):
\beqa
&&m=\Bigg|1-q_\mathrm{b}^2r^4
\Big(({x}_{1}^2+{x}_{2}^2+r^2)^2-\nonumber\\
&&-4r^2({x}_{1}\sin{\omega t}-{x}_{2}\cos{\omega t})^2\Big)\Big/
\Big(({x}_{1}^2+{x}_{2}^2+r^2)^2-\nonumber\\
&&-4r^2({x}_{1}\cos{\omega t}+{x}_{2}\sin{\omega t})^2\Big)^2
\Bigg|^{-1},
\eeqa
where $q_\mathrm{b}=\frac{8GM}{c^2}\frac{D_{ls}D_l}{D_s r^2}$
and the other values are defined as in the previous section.

Analogously to the previous case we obtain approximate formulae
for magnification
\beqa
&&m=1+\frac{q_\mathrm{b}^2 r_y^4}{\rho^4}
-\frac{4q_\mathrm{b}^3 r_y^6}{\rho^6}+\nonumber\\
&&+\frac{6q_\mathrm{b}^2 r_y^6}{\rho^8}
\big(({y}_1^2-{y}_2^2)\cos{2\omega t}
+2{y}_1{y}_2\sin{2\omega t}\big)
\eeqa
(where $r_y=r D_s/D_l$) and for the amplitude of fluctuations
in case of circular motion of the binary system:
\beq
\Delta m\approx
\frac{12 q_\mathrm{b}^2r_y^6}{\rho^6}\approx
\frac{1.2\times10^6\ \theta_{r}^{8} D_l^4 D_{ls}^2}
{c^4 T^4 \theta_I^6 D_s^2},
\label{binary}
\eeq
where $\theta_I=\rho/D_s$ and $\theta_{r}=r/D_l$ (angles in radians).
Here we have used the relation $GM=4\pi^2r^3/T^2$ for the circular motion.

\section{Explanation of the experimental data}
\label{Exp}

Finally we apply the above calculations to explain the observed brightness oscillations.
These oscillations are nearly sinusoidal, their period is approximately 100 days and their
amplitude is about $4\%$ of the quasar image brightness. At least three oscillations were
observed during the period of the observations. We consider the possibility that this
phenomenon is caused by the cosmic string loop passing through the neighborhood of images
A and B at a small distance from the observer. Obviously to fit the observational data
described in the section~\ref{Obs} we are forced to restrict the parameters of our model.
Also, because the synchronous oscillations have been observed within a limited time
interval, we  include into the consideration the motion of the loop. At that the number
of observed oscillations (3-4) restricts a transverse velocity of the loop to the values
$v_1, v_2\leq 0.1c$, but leaves a considerable freedom for the velocity component $v_3$
along the line of sight. In this case the only correction for the lens equations, as can
be shown,  is to change the parameter $q_\mathrm{s}$ by
$q_\mathrm{s}^{*}=q_\mathrm{s}(1+v_3/c)^{-1}$.

\subsection{String-induced and binary star-induced oscillations of quasar brightness}
\label{expl:osc}

\begin{figure}
\resizebox{\hsize}{!}{\includegraphics{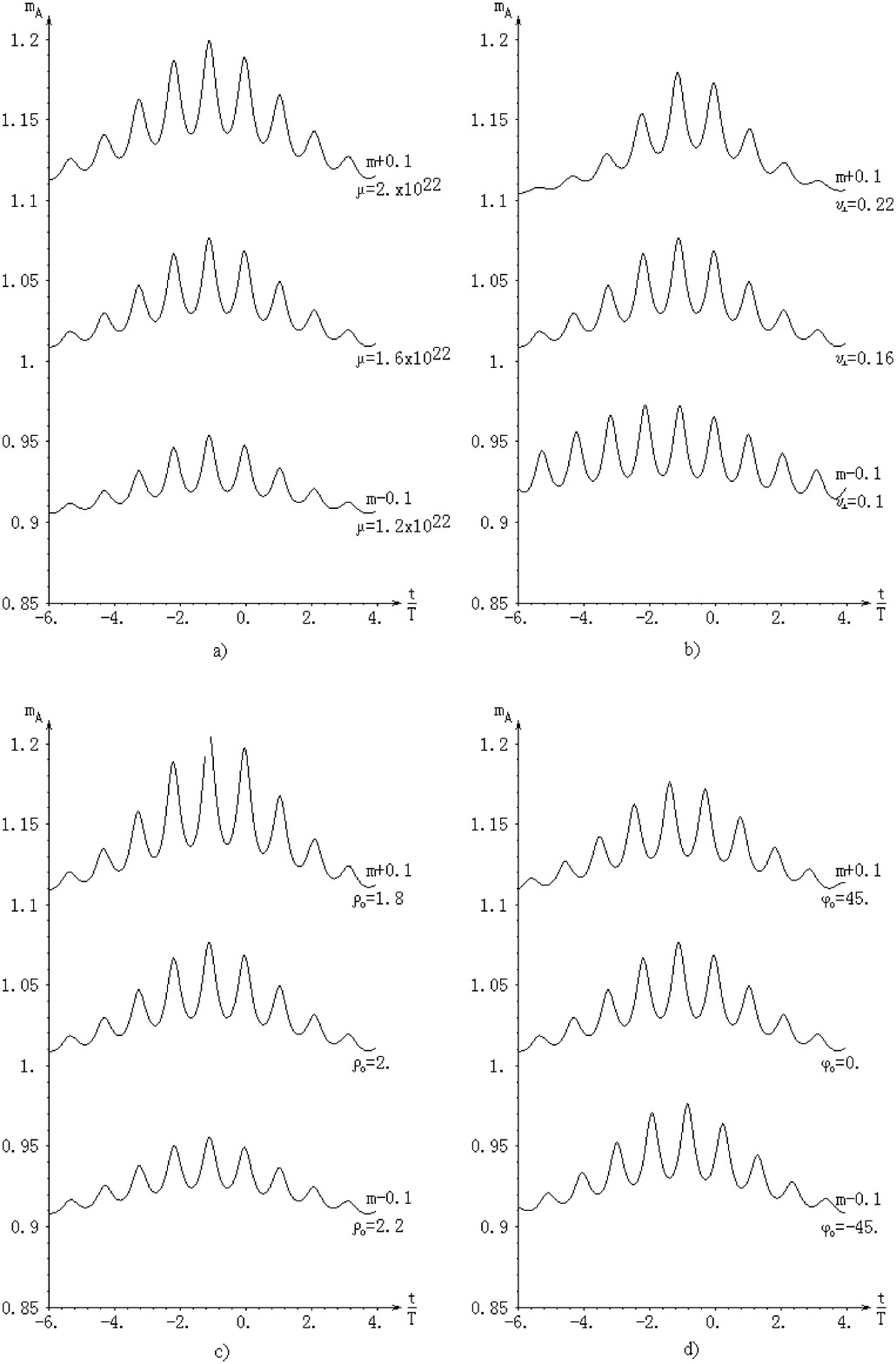}}
\caption{Oscillations of quasar image brightness caused by the loop depending:
{\bf a)} on mass per unit length $\mu$ in g/cm,
{\bf b)} on transverse velocity of loop $v_{\perp}$ in units of the speed of light,
{\bf c)} on minimal distance $\rho_0$ between image and center of the loop in units of $R$,
{\bf d)} on angle $\varphi_0$ between loop direction and direction from the loop center to
the image at $t=0$ in degrees.}
\end{figure}

As we mentioned above, the period of observed oscillations $T\approx 100$ days
is related to the string length $l=2\pi R$ as $R=cT_l/\pi$.
Taking into account relativistic motion of the string along
the line of sight we have $T_l=T(1-v_3^2/c^2)^{1/2}$.
Equation (\ref{magnif1}) can be rewritten as:
\beqa
\Delta m&\approx&5.6\bigg(\frac{\theta_I}{3\arcsec}\bigg)^{-6}
\bigg(\frac{\theta_{R}}{1.5\arcsec}\bigg)^{4}\times\nonumber\\
&&\times\bigg(\frac{\mu}{10^{22}\mathrm{g/cm}}\bigg)^{2}
\bigg(1+\frac{v_3}{c}\bigg)^{-2}.
\label{magnif}
\eeqa

\begin{figure}
\resizebox{\hsize}{!}{\includegraphics{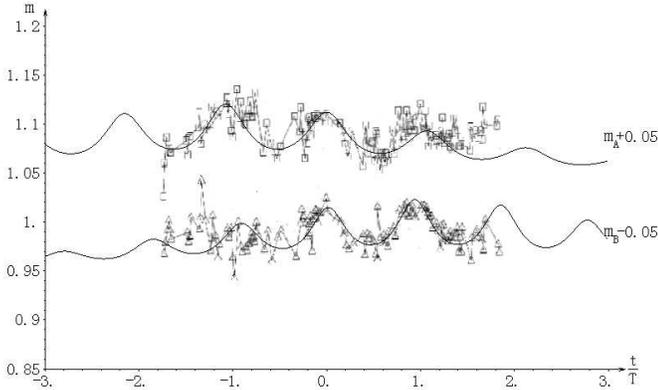}}
\caption{Oscillations of quasar image magnification predicted by the cosmic string model.
Upper and lower curves are shifted up and down by $0.05$ and fitted to image A and B
brightness records, respectively.}
\label{fluct}
\end{figure}

\begin{figure*}
\resizebox{\hsize}{!}{\includegraphics{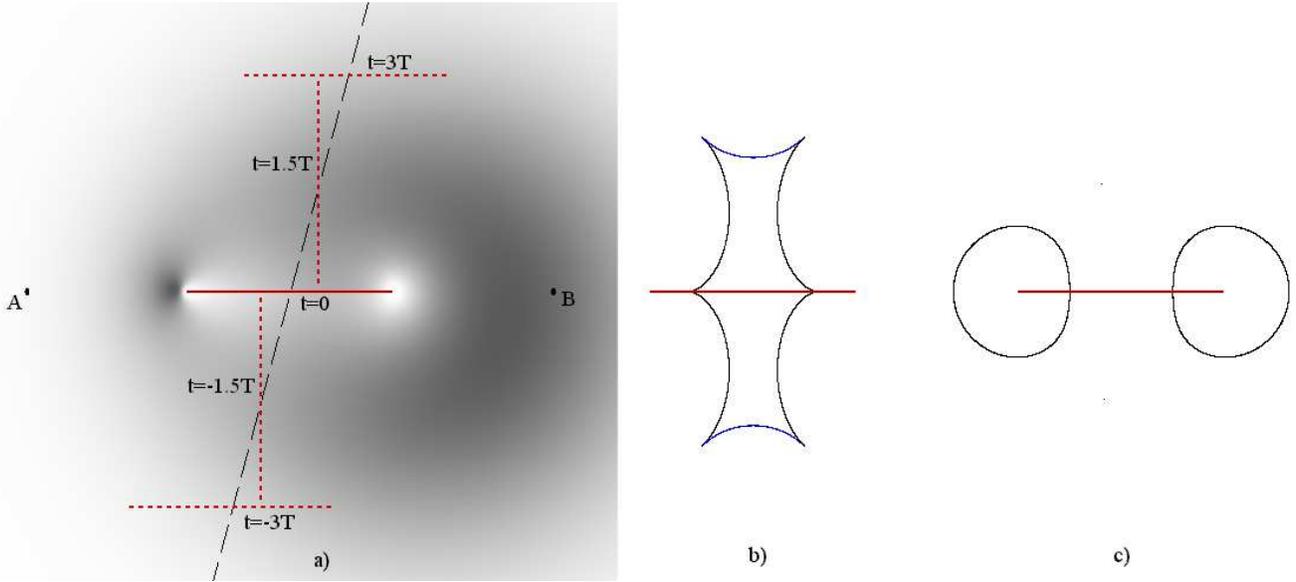}}
\caption{{\bf a)} View of the lens galaxy microlensed by the string loop at $t=0$.
A and B show the position of quasar images. Loop
positions in different moments of time are indicated as well.
Simultaneously {\bf b)} the caustics (shown sideways) and the boundaries of the image doubling
zone (upper and lower) in the plane of the galaxy and {\bf c)} the
critical curves around the loop edges in the lens plane are presented.
\label{gal&caust}}
\end{figure*}

To provide almost equal amplitude of the brightness variations in both images the loop should
fly close to the mid-point. In this case $\theta_I \approx 3\arcsec$, i.e. about half of the distance
between the images A and B. For numerical estimates we put, e. g., $v\approx 0.7c$.
In order to have 3 oscillations with possible phase shift (see Fig.~\ref{fluct}) we need
$v_1\approx 0.03$ and $v_2\approx 0.11$ (Fig.~\ref{gal&caust}).

In order to have quasi-sinusoidal variations, $\theta_R $ must be considerably smaller
than the angular distance between images A,B of the Q0957+561 (otherwise there will be
sharp spikes and/or discontinuities in the dependence of brightness upon time) and $\theta_R$
cannot be too small leading to large loop mass in virtue of equation~(\ref{magnif})
(to avoid a large monopole input into the effective lens potential leading to additional
amplification and corresponding unobserved slow increase and decrease of image brightness
superimposed on smaller oscillations due to loop rotation). Therefore we should take
$\theta_R\approx 1.5''$ and consequently $D_l\approx 3$~kpc. From equation~(\ref{magnif})
we can find that $\mu\approx 4\times10^{20}$g/cm for observed amplitude $\Delta m\approx0.04$.
More accurate values of loop parameters follow from numerical solution of
equations~(\ref{lenseq1})~-~(\ref{mag1}) without assumption $\sqrt{{y}_1^2+{y}_2^2}\gg R_y$.
For explanation of observations we need $q_\mathrm{s}=1.3$ and consequently
$\mu \approx 8\times 10^{21}$~g/cm (see Fig.~\ref{fluct}). Remarkably, this value is close to
one predicted by the GUT, $\mu \sim \mu_\mathrm{GUT}.$

In the alternative case of lensing by a binary system we obtain from equation~(\ref{binary}):
\beqa
\Delta m &\approx& 0.04
\bigg(\frac{T}{100\mathrm{~days}}\bigg)^{-4}
\bigg(\frac{\theta_r}{1.5''}\bigg)^{8}\times\nonumber\\
&&\times\bigg(\frac{\theta_I}{3''}\bigg)^{-6}
\bigg(\frac{D_l}{1.2\mathrm{~pc}}\bigg)^{4},
\eeqa
In order to explain the observed fluctuations the requirement on the trajectory should be
the same as in the loop case: $\theta_I \approx 3\arcsec$, i.e. about half of the distance
between images A and B. The minimum distance from us to this system must be $1.2$~pc,
the orbital radius should be $\theta_r D_l\approx 1.8$ a.u. and the masses of the components
should equal $78M_{\sun}$ to supply $4\%$ amplitude fluctuations of period 100 days. At such
distance the stars would be observable. For larger distances to the binary star the masses
of the components need to be larger as well. Therefore we consider such a binary star model
to be unacceptable.

\subsection{Influence of the loop on the brightness of the lensing galaxy}

Let us now consider the influence of a loop on the brightness of the lens galaxy which
is visible between the images of the quasar and makes a [ 3 \%, 15 \% ] contribution to
the observed brightness in the [ A, B ] apertures, respectively (Colley \& Schild \cite{Col99},
Fig.~3). For a galaxy at distance approximately $1\arcsec$ from image B with Gaussian
brightness distribution ($\sigma= 2\arcsec$), we obtain from numerical calculation that
relative oscillations of galaxy brightness in the measured apertures are equal to
$9\%$ and $6\%$. This corresponds to $1.6\%$ and $1.1\%$ of total signal in measured apertures
A and B, which include quasar images. Oscillations will be superimposed on a background of monopol component
with amplitudes of $12\%$ and $14\%$ of galaxy brightness. This corresponds to $2.2\%$
and $2.5\%$ of total brightness variations. A view of the lens galaxy as modified by
the loop at $t=0$ is shown in Fig.~\ref{gal&caust}~a.

The loop overlapping a galaxy can result in significant microlensing effect.
Caustics and critical curves are shown in Fig.~\ref{gal&caust} (b~and~c).
When a star of radius $R_\mathrm{s}$ is located on the caustic near its edge, the star's brightness
is increased by a factor of $3\times10^{5}(R_\mathrm{s}/R_{\sun})^{-1/2}$ according to equation for
magnification near a straight caustic (see Schneider, Ehlers, \& Falco \cite{Schneider}).
The corresponding increase of the galaxy brightness will be about $3\times 10^{-7}$ i.e.,
unobserved in our case.

The largest magnification is expected for stars crossing the cusp of the caustic.
Our calculations show that magnification of a star with $R_\mathrm{s}=R_{\sun}$, is equal to
$m\approx 1.1\times 10^{9}$ ($0.1\%$ relative to the galaxy brightness).
The visible speed of the cusp motion through  the galaxy plane is
$v_\mathrm{cusp}\approx 2\times 10^{-3}$~pc/s. In one pixel of a telescope image plane,
light is collected from about
\mbox{$(D_\mathrm{g}\theta_\mathrm{pixel})^{2}n_\mathrm{star}\approx5\times10^8(\theta_\mathrm{pixel}/0.1'')^2$}
stars (\mbox{$n_\mathrm{star}\approx1000\mathrm{~pc}^{-2}$} is the surface density of
stars, \mbox{$D_\mathrm{g}\approx1.4$~Gpc} is the distance to the galaxy).
Therefore during about $10^{-8}(\theta_\mathrm{pixel}/0.1'')^{-3}\approx4\times 10^{-11}$~s
the brightness of the star is larger than the brightness of the pixel without lensing.

For telescope integration time $T_\mathrm{int}$, the average magnification of the star
in the cusp is: \mbox{$<m>=1.2\times 10^{4}(T_\mathrm{int}/1\mathrm{s})^{-2/3}\approx400$}.
The typical distance between projections of stars in the galaxy plane equals
$l_\mathrm{ss}\approx0.03$~pc. Therefore such flashes will be repeated approximately
every $l_\mathrm{ss}/v_\mathrm{cusp}\sim10$ seconds. In our case the galaxy brightness is
about $20\%$ of the quasar B image brightness and therefore the maximum of these flashes
will be approximately $2\times10^{-4}$ of the image brightness, which is below observational
precision. So lensing of the galaxy is observable only from distortion of the total galaxy
image, as shown in Fig.~\ref{gal&caust}~a.

\section{Conclusions}
\label{Dis}

Our main motivation to apply the  cosmic string model for explanation of brightness
oscillations in question is their specific characteristics. It is difficult to propose
a less exotic model, such as a double star model to explain the observed oscillations.
We have shown (see subsection \ref{expl:osc}) that in case of a double star the masses
of the components must be of  order of $78M_{\sun}$ at $1.2$~pc distance in order to meet
all observational requirements. For larger distances the masses of the components should
be larger as well. This seems to be unacceptable.

On the other hand the property of fast oscillations is typical for cosmic strings.
To show viability of the string interpretation of observed oscillations we have chosen
a particular ``degenerate'' (highly asymmetrical) analytical solution of string equations.
This solution is a limiting case of strongly elongated rotating loop configuration with
sufficient quadrupole moment. Of course, more realistic case should include additional
modes of loop oscillations in order to avoid  self-intersections and annihilation.
Nevertheless, even in more general cases solutions with sufficient quadrupole moment can
provide the same level of brigthness variations.

The results presented here show that loops of cosmic strings supply quantitative explanations
of synchronous variations in the two images of the gravitationally lensed quasar Q0957+561
A,B. The derived value of the string parameter $\mu\sim 10^{22}$g/cm lies just in the
theoretically preferable range. Atypical (with small probability of realization, but not
impossible) in our model are the distances to the loops and their sizes - both are about
$10^{-3}-10^{-4}$ of statistically expected values. The reason for this is the  relatively
short observed period, only of order 100 days, which limits the length of the loop. Moreover,
the observed angular separation of the images predicts the distance to the loop, while we
fixed its size according to the period of flux fluctuation. Consequently in another
hypothetical object with different observational parameters, more typical loops will probably
work. Of course, a single event need not follow statistical rules, especially since our
observational sampling rate might strongly favor this particular specimen. Moreover, some
physical mechanism might cause the concentration of loops in galactic halos. Therefore,
further observational efforts towards uncovering more objects with similar properties are
extremely important.

Searches for brightness oscillations, similar to those described above, can be a promising
way of discovering the gravitational signatures of cosmic strings.

\acknowledgements{We are indebted to anonymous referee for useful comments. We thank Patrick B. M. van Kooten
for a helpful reading of this manuscript.}


\begin{thebibliography}{}

\bibitem[2003]{Col03} Colley, W. et al. 2003, ApJ, 587, 71

\bibitem[1999]{Col99} Colley, W. \& Schild, R. 1999, ApJ, 518, 153

\bibitem[2003]{CS03} Colley, W. \& Schild, R. 2003, ApJ, 594, 97

\bibitem[1996]{Laix} de Laix, A. \& Vachaspati, T. 1996, Phys. Rev. D, 54, 4780

\bibitem[2003]{Ova03} Ovaldsen, J. et al. 2003, A\&A, 402, 891

\bibitem[1964]{Ref64} Refsdal, S. 1964, MNRAS, 128, 307

\bibitem[2003]{Sazetal03} Sazhin, M. et al. 2003, MNRAS, 343, 353

\bibitem[1999]{RS99} Schild, R. 1999, ApJ, 514, 598

\bibitem[1986]{SC86} Schild, R. \& Cholfin, B. 1986, ApJ, 300, 209

\bibitem[1991]{SS91} Schild, R. \& Smith, R. C. 1991, AJ, 101, 813

\bibitem[1995]{ST95} Schild, R. \& Thomson, D. 1995, ApJ, 109, 1970

\bibitem[1992]{Schneider} Schneider, P., Ehlers, J. \& Falco, E. E. 1992,
Gravitational Lenses (Berlin: Springer-Verlag)

\bibitem[1984]{Turok} Turok, N. 1984, Nuclear Physics B242, 520

\bibitem[1994]{VilShe94} Vilenkin, A. \& Shellard, E. P. S. 1994, Cosmic Strings and Other Topological Defects (Cambridge: Cambridge University Press)

\bibitem[1979]{Wal} Walsh, D., Carswell, R., \& Weymann, R. 1979, \nat, 279, 381

\end{thebibliography}
\end{document}